**One-Dimensional Twisted and Tubular Structures of Zinc Oxide by Semiconductor-Catalyzed Vapor-Liquid-Solid Synthesis**

Thang Pham[1,2], Sampath Kommandur[2], Haeyeon Lee[1], Dmitri Zakharov[3], Michael A. Filler[2] and Frances M. Ross[1]*

[1]Department of Materials Science and Engineering, Massachusetts Institute of Technology, Cambridge, MA 02139, USA

[2]School of Chemical & Biomolecular Engineering, Georgia Institute of Technology, Atlanta, GA 30332, USA

[3]Center for Functional Nanomaterials, Brookhaven National Laboratory, Upton, NY 11973, USA

Email: fmross@mit.edu

**Abstract**

The exploration of new catalysts for the vapor-liquid-solid (VLS) synthesis of one-dimensional (1-D) materials promises to yield new morphologies and functionality. Here, we show, for the model ZnO system, that this possible using a semiconductor (Ge) catalyst. In particular, two unusual morphologies are described: twisted nanowires and twisted nanotubes, in addition to the usual straight nanowires. The twisted nanotubes show large hollow cores and surprisingly high twisting rates (up to $9°/\mu m$), which cannot be easily explained through the Eshelby twist model. A combination of *ex situ* and *in situ* transmission electron microscopy measurements suggest that the hollow core results from a competition between growth and etching at the Ge-ZnO interface during synthesis. The twisting rate is consistent with a softening of elastic rigidity. These results indicate that the use of unconventional, nonmetallic catalysts provide opportunities to synthesize unusual oxide structures with potentially useful properties.

Supplementary material for this article is available online.

Keywords: VLS, nanotubes, ZnO, germanium catalyst, Eshelby twist

## 1. Introduction

One-dimensional (1-D) metal oxide nanostructures with different morphologies have been synthesized via the vapor-liquid-solid (VLS) mechanism using metal catalysts [1]. Gold (Au) is the most common choice. However, the use of any metal as a catalyst presents two problems for metal oxide nanostructure growth. Metal catalysts, especially Au, are known to dope the nanostructure or diffuse into it during growth [2,3], which may create mid-gap non-radiative levels and hence degrade the physical properties. Other catalyst metals may have a strong interaction with the metals that are of most interest in complex oxides, which may hinder growth since the synthesis process requires a catalyst that can act as a reservoir for all the metals involved. These issues have led to an active search for alternative catalysts for the VLS synthesis of oxide nanowires (NWs) [4]. Recent studies demonstrate that a semiconductor, germanium (Ge), can act as an effective catalyst for growth of zinc oxide (ZnO) NWs [5,6]. Ge is an intrinsically interesting catalyst for oxide NWs because it is easily oxidized to form volatile species such as GeO$_2$, a process that we might expect would keep the liquid surface open to accommodate the continuously arriving metal precursors. The use of Ge catalysts may therefore broaden the synthesis space and the opportunities to create new types of metal oxide 1-D nanostructures. Compared to conventional metal catalysts (e.g., Au), Ge catalysts are unusual in showing an extremely large contact angle at the vapor-liquid-solid interface [5,6]. The reason for this effect and its influence on growth is not known.

Here, we examine how a Ge catalyst can facilitate the formation of unusual structures in metal oxides. We are particularly interested in synthesis of nanostructures that are hollow and/or twisted due to their intriguing growth modes [7,8] and properties [9–11]. Hollow nanotubes (NTs) have been synthesized in polycrystalline form, for example by using sacrificial templates [12] or by exploiting the Kirkendall effect [13]. Single crystal NTs, especially those with a periodic spiraling structure, are mostly found in layered materials [8], including chiral carbon and MoS$_2$. Tubular and twisted structures in non-layered materials often form in the presence of an axial screw dislocation [14]. The strain field created around the disrupted region causes the crystal to twist around its axis, a phenomenon known as the Eshelby twist [15]. The presence of the dislocation may also drive another effect, the Frank open-core mechanism [16,17], in which a hollow core is energetically favorable whenever eliminating the highly strained material at the dislocation core overcomes the energetic cost of introducing new inner surfaces. The coexistence of both mechanisms creates NTs that are both hollow and twisted [18] with a twisting rate of below 2$^\circ$/μm [17].

We will show that the complex behaviors of Ge catalysts may produce unusual structures in metal oxides. We combine *ex situ* and *in situ* electron microscopy and spectroscopy to quantify the prevalence of each type of structure as a function of growth conditions, and to determine structural properties such as catalyst geometry, diameter, facet structure, and twisting rate. We find that the catalyst geometry is similar for all morphologies and is consistent with a VLS mechanism. However, we find unusual features in the twisted nanostructures. First, the cores are much larger than expected from the hollow-pipe mechanism. And second, the twisted NTs show much higher twisting rates than can be explained by Eshelby twist, even though Eshelby's model does account for the twisting rate in the twisted solid NWs observed here. We will discuss how features of the Ge-catalyzed growth mechanism may explain both the large cores and the high twisting rates in the NTs. Based on *in situ* transmission electron microscopy (TEM) movies, we suggest that there is a competition between growth and etching processes at the catalyst/nanowire interface. We propose that etching is responsible for the large hollow cores and we suggest that the NT twisting anomaly can be explained by changes in mechanical properties due to catalyst diffusion under the growth conditions. We suggest that nanostructure formation processes in which there is a competition between growth and catalyst-induced etching may be useful for controlling structure and properties, and may lead to opportunities that are unavailable from VLS growth using conventional metal catalysts. The results emphasize the importance of exploring new catalysts as a strategy to gain more precise control over oxide nanowire growth.

## 2. Methods

*2.1. Materials synthesis*

Ge-catalyzed ZnO materials were grown by thermal vapor transport using zinc oxide, black carbon and germanium oxide powders as precursors (1:1:0.1 molar ratio). The mixed precursors (0.5 gram) were put at the central of a horizontal tube furnace, which was set at 975-1000 °C. The growing substrates, which included silicon wafers, silicon nitride coated wafers and silicon nitride TEM grids, were put downstream, 10-12 cm away from the precursors. The estimated temperature at the substrates is 500-650 °C. Each growth is typically 15-30 minutes long, under a flow of 50-65 sccm $N_2$/Ar as the carrier gas. After growth finishes, the furnace cap is opened and the carrier gas flow is changes to 200 sccm to facilitate quick cooling. We do not observe any significant difference in material morphologies, distribution and yield when using different substrates.

*2.2. Ex-situ electron microscopy characterization*

SEM images and SEM-EDS spectroscopic data were obtained for nanostructures without removing them from the growth substrate, using a FEI FIB/SEM working at 5-20 kV. TEM samples were prepared on copper grids by either dry or wet transfer. For dry transfer, copper grids were gently dragged over the growth substrates to collect the grown material. For wet transfer, the growth substrates were sonicated briefly in iso-propanol solution and the solution was drop-cast onto copper grids. Imaging and selected area diffraction were carried out on a JEOL 2010F TEM operated at 200 keV. Diffraction patterns of ZnO for different zone axes were simulated using Crystal Maker. We note that because of their large sizes, Ge catalysts were easily broken off during the transfer processes. As a result, many TEM images in this study show only ZnO stems.

*2.3. In-situ environmental TEM*

Ge-catalyzed ZnO nanostructures were grown directly onto SiN TEM grids. The grids have several opened windows where ZnO can grow off the SiN. This configuration permitted observation of Ge-ZnO interactions *in situ* using the aberration-corrected Thermo Fisher Titan 80-300 environmental TEM [19] at Brookhaven National Laboratory, operated at 300 keV. Samples were heated using a Gatan double tilt heating holder and digermane (20% in He) was flowed into the sample area using a manual leak valve to a total pressure of $1$-$2 \times 10^{-5}$ Torr. Note that the ETEM base pressure is $10^{-6}$ Torr. The electron dose rate ($10^3$-$10^6$ electron/nm$^2$s) was monitored to minimize its effect on the observed phenomenon. The electron dose for data presented in figure 4 are specified in the corresponding movies in the supplementary data.

*2.4. STEM-Cathodoluminescence*

STEM-CL images and spectra were collected in a JEOL 2011 TEM connected to a Gatan MonoCL3+ CL system with photomultiplier tube detector. The setup allows the collection of both CL maps in panchromatic/monochromatic modes and point spectra. Panchromatic CL map collects the light emission with a full range of wavelengths.

**3. Results and Discussion**

*3.1. Morphologies of Ge-Catalyzed ZnO Nanomaterials*

Ge-catalyzed ZnO nanostructures were synthesized by a vapor transport method using zinc oxide, carbon black, and germanium oxide powders as precursors [5]. Details of the procedure are given in the Methods. A significant feature of the growth process is that Ge must be supplied continuously for growth to be sustained, as has been noted previously.

Figure 1 displays an overview of the range of ZnO nanostructures that result from this synthesis. We observed three distinct morphologies: straight NWs, twisted NWs, and twisted NTs. Unless broken, each morphology exhibits a large, spherical catalyst particle at the tip. Energy dispersive spectroscopy (figure S1, supplementary data) confirms that all morphologies are composed of ZnO and the catalyst particle is Ge. TEM images and diffraction patterns, shown for a representative NT in figure 1(d), reveal that both NWs and NTs have single crystalline wurtzite structure growing in the [0001] (c-axis) direction. Diameters range from 100 to 2000 nm and lengths can be up to 50 μm after 30 minutes of growth. The Ge catalysts have a diameter between 1 and 8 μm, typically 4 times the ZnO diameter for all morphologies [6] (figure 1(e-f)). The growth rate is 6-20 nm/s, which is comparable to previous reports using Ge catalyst [5,6] (table S1, supplementary data) and 5-10 times faster than Au-catalyzed ZnO NWs synthesized under similar conditions [4,20]. The percentage of structures with each morphology depends on the growth conditions (figure S3, supplementary data). At the cooler end of the furnace (500-575 °C), solid NWs dominate (> 80 %) with twisted NWs and NTs being much less common (5-10 % each). In contrast, twisted NTs are more frequently observed (30 %) at the hotter end of the furnace (575-650 °C). The observation of hollow and twisted structures (figure 1(b-c) suggests that axial screw dislocations are present [17,21–24]. Indeed, diffraction-contrast TEM measurements confirm the existence of dislocations (figure S2, supplementary data).

Although these nanostructures appear to broadly follow the VLS growth mechanism, the diversity of hollow and twisted structures indicate that the growth may include aspects that are not part of the general VLS model [5,6,20,25]. We therefore analyze in detail the twisted NTs and NWs to gain additional insight into their formation process.

*3.2. Motif 1: Hollow cores*

The cleaved ends of NTs show that these structures have a range of internal diameters, as displayed in figure 2(a) and 2(b). Furthermore, we observe fully or partially hollow structures, and occasional structures that include a Ge particle within the cavity (figure S4, supplementary data). Figures 2(c-d) show that there is no strong relationship between the inner (r) and outer (R) diameters, other than the obvious requirement that r < R. Values of r/R range from 0.34 to 0.75 with a mean of 0.55.

These hollow structures are also twisted, as will be discussed in the following section. The presence of hollow cores in these NTs, especially given their twist, is immediately reminiscent of the open channels that are often associated with screw dislocations [17]. Such open cores are commonly explained by Frank's elastic open-core mechanism [16,17]. The model states that, in order to relieve the strain energy at the dislocated region, the disrupted core is removed at the penalty of introducing new surfaces (the surfaces of the hollow cavity). Minimization of energy under these circumstances leads to a 1-D hollow structure with inner diameter given by equation (1):

$$r_{Frank} = \left(\frac{\mu}{8\pi^2\gamma}\right)b^2 , \qquad (1)$$

where b is the magnitude of the Burgers vector, μ is the shear modulus, and γ is the surface energy. For ZnO with $b_{ZnO,NT}$ = ~2.0 nm [17], $\mu_{ZnO}$ = 51.0 GPa [26], $\gamma_{ZnO}$ = 0.31 J.m$^{-2}$ (Ref. [17]), the corresponding Frank's hollow radius is $r_{Frank}$ = 26 nm. This is smaller than any of the inner radii observed for our NTs: Figure 2(c) shows $r_{NT}$ in the range 50-500 nm, which is 2-6 times larger than predicted by Frank's model. We will discuss the implications of these anomalously large cores in the following sections.

*3.3. Motif 2: Axial twisting*

The second striking feature of these Ge-catalyzed ZnO nanostructures is the twisting seen in NTs and some of the NWs. TEM provides a direct means of measuring the twisting rate in both NTs and NWs. As shown in figure 3(a-b), selected-area diffraction patterns are collected at different locations along a nanostructure and matched to simulations with the orientation of the crystal at that location. The presence of different and alternating crystallographic orientations in the nanostructure is proof of its continuously twisting structure [21–24]. For the NT shown in figure 3(a), we determine a twisting rate of 8.7 °/μm, more than 4 times greater than that measured (< 2 °/μm) for smaller diameter ZnO NTs synthesized in solution [17]. Additional examples of similar twisting rate measurements are shown in figure S5 (supplementary data).

For 1-D materials having an axial screw dislocation, Eshelby proposed that a pathway to compensate the strain energy in the dislocation core is to twist the entire structure. In the framework of continuum elasticity, the Eshelby twist is equivalent to an application of a torque, *T*, exerted at two ends of a solid wire with a length, *L*, that results in a torsional moment $M_{Eshelby} = TL$ [15]. The Eshelby induced moment leads to a twisting angle $\alpha$ (per unit length) of:

$$\alpha_{Wire} = \frac{M_{Eshelby}}{\mu_{Wire} \cdot J_{Wire}} = \frac{M_{Eshelby}}{\mu_{Wire} \cdot \frac{\pi}{2} R^4} \qquad (2)$$

where $J_{Wire}$ is the polar moment of inertia of a solid wire, $J_{Wire} = \frac{\pi}{2} R^4$, and $\mu_{Wire}$ is the modulus of rigidity (shear modulus) of the wire. According to Eshelby [15], the torsion moment $M_{Eshelby}$ is the result of the energy balancing with the strain energy stored in the screw dislocation, therefore $M_{Eshelby} = \frac{1}{2} \mu_{Wire} b R^2$. Consequently, the Eshelby twisting rate $\alpha_{Wire}$ (in units rad/μm) is:

$$\alpha_{Wire} = \frac{\frac{1}{2} \mu_{Wire} \cdot b \cdot R^2}{\mu_{Wire} \cdot \frac{\pi}{2} R^4} = \frac{b}{\pi R^2} \; . \qquad (3)$$

We plot the twisting rate α as a function of $1/\pi R^2$ in figure 3(c). Twisted NWs exhibit twisting rates consistent with equation (3). The slope of the linear fit yields a Burgers vector $b_{Twisted\text{-}Wire}$ = 2.3 nm, approximately 4 times the lattice parameter in the [0001] orientation ($c_{ZnO}$ = 0.53 nm) and of the same order as that seen in other NWs (0.5-3.0 nm) [17,18,21–24]. However, twisted NTs do not follow the same relationship. Even if we include the hollow core by including the appropriate geometric correction to the polar moment of inertia [17,27], the best fitting line yields a much larger, and likely physically unrealistic, Burgers vector of $b_{Twisted\text{-}Tube}$ = 5.4 nm. We discuss the anomalously large twisting rate in NTs in the next sections.

*3.4. Ge-ZnO Interaction: Growth vs. Etching*

Our *ex situ* observations reveal shared features among all Ge-catalyzed ZnO nanostructure morphologies, such as the overall dimensions and relative size of the catalyst compared to the diameter, suggesting commonalities in the catalytic growth mechanism. Furthermore, the twisted NWs show twisting rates that are consistent with physically reasonable Burgers vectors. However, three features of the Ge-catalyzed ZnO nanostructure growth require additional explanation. First, Ge must be supplied continuously if growth is to be sustained. Second, the diameter of the core (and even its presence) can vary along the length of the NT and the dimensions of the hollow cores can be several times larger than expected from Frank's micropipe model. There is also a striking difference in the degree of faceting of the inner and outer walls. Third, the NTs show twisting rates that are several times larger than expected from Eshelby twist, even accounting for the hollow geometry. These anomalies suggest that ZnO growth using Ge catalysts is more complex than the VLS process that has been described in the case of metal catalyst counterparts.

We believe that the mobility and activity of Ge during growth can explain all the anomalous features of growth. The requirement for continuously supplying Ge suggests that Ge atoms leave the catalyst during growth, either by surface diffusion or (perhaps more likely at high temperature) by oxidation and evaporation via the formation of volatile species, such as $GeO_2$. These dynamics, in addition to the presence of Ge within the NT cavity (figure S4, supplementary data), hint at potential interactions between Ge and ZnO at elevated temperatures.

In order to assess any such interactions directly, we examine the behavior of Ge and ZnO NWs via *in situ* environmental TEM (ETEM). Figure 4 shows a series of images obtained during heating under vacuum of NWs that were still attached to their catalysts. Visible changes started to occur as the temperature was raised to 650 °C. A gap appeared at the Ge/ZnO interface, as displayed in figure 4(a), suggesting an etching reaction between Ge and ZnO. Another example is shown in figure 4(b) where residue of the catalyst particle created a pit in the body of the NW. The etching process can also take place in the interior of a nanowire. Figure 4(c) (see also movie 1, supplementary data) shows a Ge-containing particle within a hollow core. As etching continues, the core diameter increases and the particle is eventually consumed.

To investigate the reaction under more controlled conditions, we examined the effect of adding Ge to the surface of the pre-grown ZnO NWs. This was achieved by flowing digermane ($Ge_2H_6$), which cracks to deposit Ge on the NW surface as shown in figure 4(d) (see also figure S6, supplementary data and Experimental Section for the deposition procedure). In figure 4(d), we see that these Ge particles can also etch the NW to leave a porous structure with interconnected etched pockets. The etching process is a solid-state reaction and, as shown in figure 4(e), is anisotropic with faceted etched regions visible (see also movie 2, supplementary data). As the process continues, the particles are consumed and the etching reaction eventually stops.

Figure 4(f-g) are two examples of atomic-resolution TEM images displaying the Ge-assisted ZnO etching process (see also movie 3, supplementary data). The reaction removes material at the area of contact, leaving a trace of empty volume behind that shows reduced contrast in the TEM image. The inset of figure 4(g) shows that the Ge-containing particle remains crystalline during the etching process. The fastest etching rate is measured in the [0001] direction, with a rate of 5-7 nm/s (table S1, supplementary data). Recall that the rate of Ge-catalyzed ZnO growth is between 6 and 20 nm/s, also along the [0001] orientation. Thus, the rates of etching and growth for the Ge-assisted

ZnO system are of the same order and take place over the same temperature range (600-650 °C).

*3.5. Proposed Mechanism of Anomalous Motifs*

The *in situ* observations imply a coexistence of growth and etching reactions during the formation of ZnO nanostructures from Ge catalysts. We suggest, based in particular on movie 2 (supplementary data), that the formation of unexpectedly large hollow cores takes place as etching occurs along the core of a dislocation that is already present in a solid NW. Etching processes are well known to occur more rapidly at locations with imperfect bonding such as the strained cores of dislocations [28]. The reactive Ge may even move along a narrow cavity first created by Frank's micropipe mechanism. Accelerated etching at the center of the NW would result in a large central hollow core, and the facets may result from anisotropy in the etching process.

The transport of Ge into the NT interior may also explain the anomalous twisting of the NTs. We first note that if the hollow cores are caused by etching, the Burgers vector of the screw dislocation is likely to be the same as that of the unetched NWs, *i.e.* 2.3 nm (figure 3(c)). Without any external forces, the only mechanical application on the NT is the Eshelby induced moment $M_{Eshelby}$, which is unchanged and stored within the twisted structure before the etching event. The resultant twisting rate for the NT is

$$\alpha_{Tube} = \frac{M_{Eshelby}}{\mu_{Tube} \cdot J_{Tube}} = \frac{\frac{1}{2}\mu_{Wire} \cdot b \cdot R^2}{\mu_{Tube} \cdot \frac{\pi}{2}(R^4 - r^4)} \qquad (4)$$

where $J_{Tube}$ is the polar moment of inertia of a hollow tube, $J_{Tube} = \frac{\pi}{2}(R^4 - r^4)$, and $\mu_{Tube}$ is the modulus of rigidity. Equation (4) can be rewritten as:

$$\alpha_{Tube} = b \cdot \left(\frac{\mu_{Wire}}{\mu_{Tube}}\right) \cdot \left(\frac{R^2}{\pi(R^4 - r^4)}\right) = b \cdot K_{Rigidity} \cdot K_{Geometry} \qquad (5)$$

where $b = 2.3$ nm (figure 3(c)), $K_{Rigidity} = \left(\frac{\mu_{Wire}}{\mu_{Tube}}\right)$ is the ratio of moduli of rigidity (or shear moduli) of as-synthesized NWs and etch-induced NTs, and $K_{Geometry} = \left(\frac{R^2}{\pi(R^4 - r^4)}\right)$ is the geometry corrected factor. $K_{Geometry}$ is known, based on measured outer and inner radii of the NTs (R and r). Thus on the right hand side of equation (5), the only unknown variable is $K_{Rigidity}$.

We plot the calculated $\alpha_{Tube}$ as a function of $K_{Geometry}$ for several values of $K_{Rigidity}$ in figure 5. The measured twist angles of the NTs (the data shown in figure 3(c)) are most consistent with $K_{Rigidity} = 3 - 4$, and the possibility that the anomalous twisting results from a Ge-induced softening of the ZnO in NTs. The close proximity of Ge and ZnO in the NT core may facilitate the diffusion of Ge atoms into the ZnO (figure S4, supplementary data). Several reports have in fact shown dopant-induced reduction of elastic moduli [29–31], including reductions in the Young's and shear moduli in ZnO NWs by a factor of 2. More importantly, Ge etching may induce defects such as zinc or oxygen vacancies. These defects are reported as the main sources of reduced mechanical stiffness of ZnO NWs [32,33]. Indeed, nano-probe cathodoluminescence (CL) in conjunction with scanning TEM (STEM) confirms a high concentration of the defects in the twisted NTs (figure S7, supplementary data). Additionally, a comparison [34] of the mechanical properties of ZnO NWs synthesized by electrodeposition and hollow NTs prepared via post-synthetic solution *etching* of solid ZnO NWs [34,35] showed a 5-fold higher elastic stability in ZnO NWs compared to the etch-induced ZnO NTs. This behavior is consistent with the hypothesis that the anomalous twisting in Ge-etched induced ZnO NTs originates from the reduction of the modulus of rigidity.

**4. Conclusion**

We have shown the roles of an unconventional, non-metal catalyst in controlling metal oxide morphologies, with Ge-catalyzed growth of 1-D ZnO as a model system. The VLS synthesis results in a variety of intriguing structures including straight NWs, twisted NWs, and twisted NTs, as characterized by electron microscopy and spectroscopy. *In situ* TEM reveals a competition between etching and growth at the Ge-ZnO interface. We propose that this mixed growth and etching process can explain the formation of hollow core NTs, and we speculate that the presence of Ge within these cores may explain the anomalously large twisting rate. We believe that the observed phenomenon, where there is a coexistence of material addition and removal at the interface, is quite general, and could be realized in other metal oxide nanosystems by an appropriate choice of catalysts.

The structure and properties of highly twisted structures, especially of non-layered materials, is fundamentally intriguing. For instance, the twisted NTs that exhibit inner facets and a smooth outer wall may serve as an ultraviolet lasing medium [36] with optical features distinct from those of ZnO NWs with hexagonal or rectangular cross-sections. The large built-in torsional strain of the twisted ZnO NWs and NTs may lead to enhanced piezoelectric properties [37]. In addition, the possibility exists to create a homojuction between a twisted NW and twisted NT by choice of synthesis parameters, perhaps yielding new properties and potentially useful applications.

These results underscore the critical importance of testing new catalysts for controlling 1-D oxide nanostructures via VLS, establish a foundation to explore various pathways for oxide NWs and NTs synthesis, and emphasize the importance of *in situ* measurements to investigate these structures.

**Supplementary data**

Supplementary material for this article is available online.

**Acknowledgements**

We acknowledge funding from the U.S. Department of Energy, Office of Basic Energy Sciences, Division of Materials Sciences and Engineering under Award DE-SC0019336. This work made use of the Shared Experimental Facilities supported in part by the MRSEC program of the National Science Foundation under award number DMR-1419807; and the Center for Functional Nanomaterials, Brookhaven National Laboratory supported by the U.S. Department of Energy, Office of Basic Energy Sciences, under contract DE-AC02-98CH10886. The authors thank Prof. Jeehwan Kim for assistance with material synthesis, Dr. Baoming Wang for helping with sample preparation, and Kate Reidy for fruitful discussions.

**Conflict of Interest**

The authors declare no conflict of interest.

**Figures**

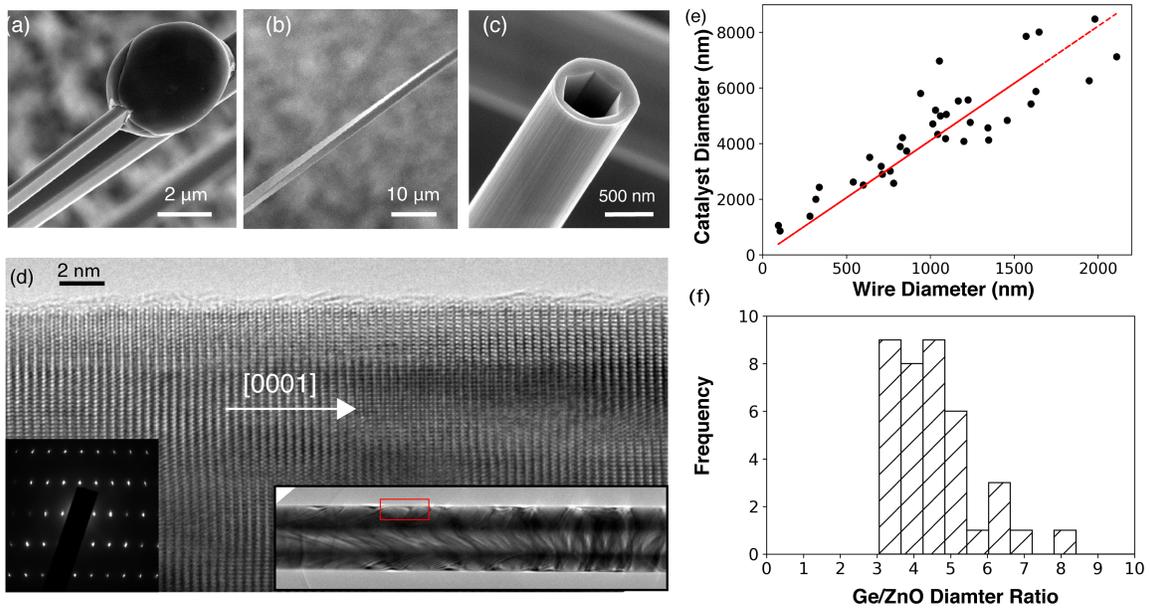

**Figure 1.** Ge-catalyzed ZnO nanostructures. (a) SEM image of a straight NW with large catalyst at the tip. (b) SEM image of a twisted NW where the rotation reveals different facets. (c) SEM image of a broken NT showing its internal structure with faceting on the inner surface. (d) High-resolution TEM (HR-TEM) image and corresponding diffraction pattern of a ZnO NT. The inset shows a low-magnification image of the same NT. The red box indicates the area where the HR-TEM was obtained. (e) Correlation between diameters of NW and Ge particle for straight ZnO NWs. (f) Distribution of Ge and ZnO diameter ratios showing that the catalyst is on average 4x larger than the nanostructures.

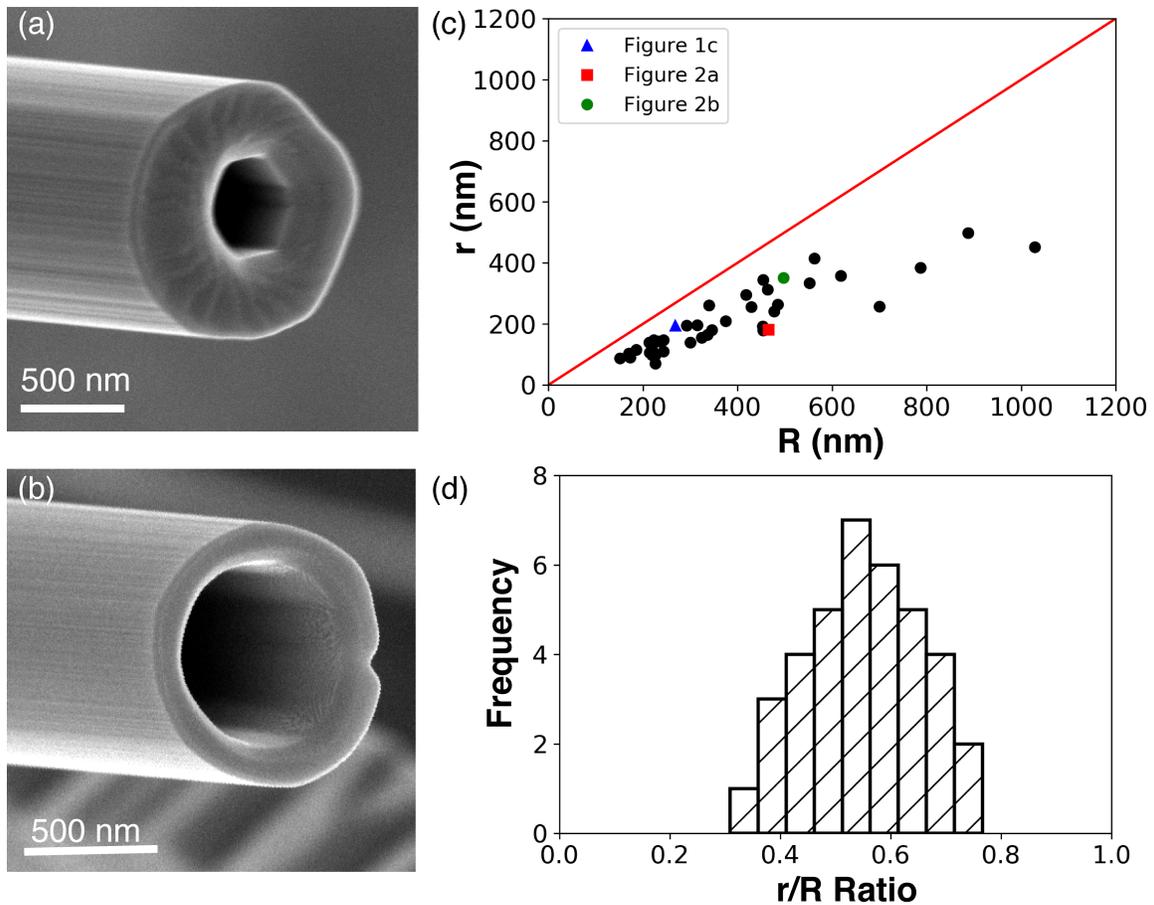

**Figure 2.** Hollow structures. (a-b) SEM images of NTs with thicker and thinner walls, respectively. (c) Relationship between NT inner and outer diameters r, R. The red line indicates r = R. The NTs shown in figure 1(c), 2(a) and 2(b) are marked. (d) Distribution of ratios between inner and outer radii.

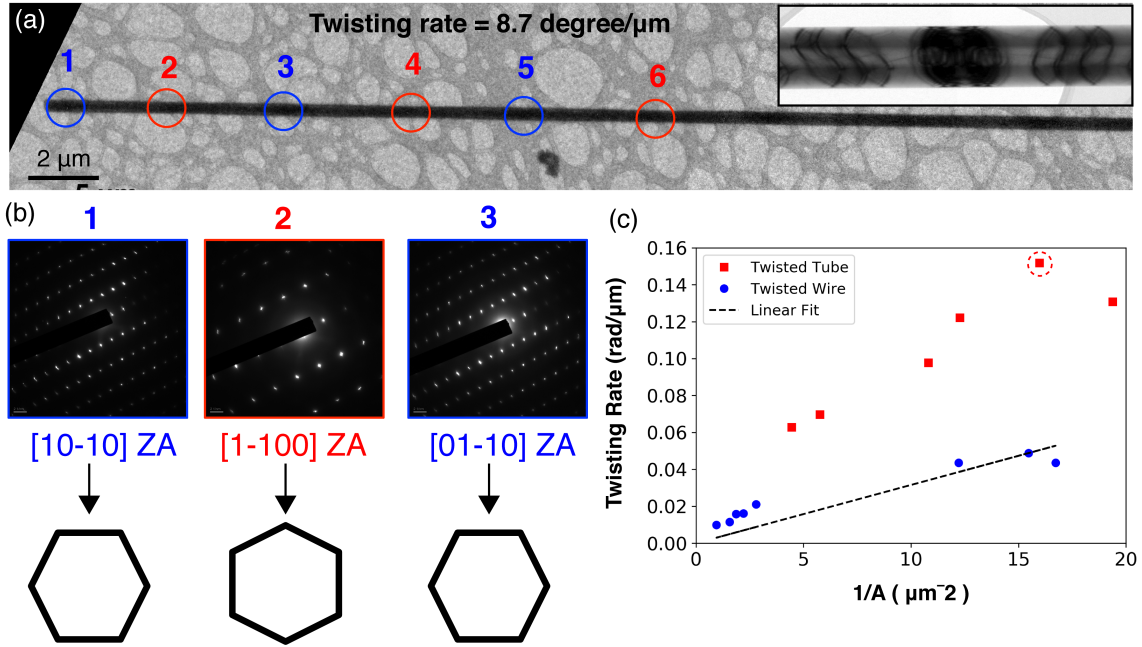

**Figure 3.** 1-D torsional structures of ZnO. (a) Low magnification TEM bright field image of a twisted ZnO NT having R = 155 nm and r = 65 nm. Areas marked as 1, 3 and 5 in blue have the same crystallographic orientation, the [10-10] ZA, while 2, 4, and 6 in red have [1-100] ZA, as determined from the selected-area diffraction patterns in (b). The corresponding simplified schematics are illustrated below. The crystal rotates by 30º between each area leading to a measured twisting rate of 8.7 º/μm. (Note that the TEM image in (a) is rotated 25º clockwise with respect to the patterns in (b).) The inset of figure 3(a) is a higher magnification TEM image of the NT showing defect contrast that is consistent with twisting. (c) Twisting rate (radian/μm) as a function of inverse cross sectional area $1/\pi R^2$ for twisted NWs (blue dots) and twisted NTs (red squares) measured by TEM. The NT shown in figure 3(a) and 3(b) is marked by a dotted circle. The linear fit for twisted NWs is consistent with the Eshelby model and yields a Burgers vector of $b_{Twisted-Wire}$ = 2.3 nm.

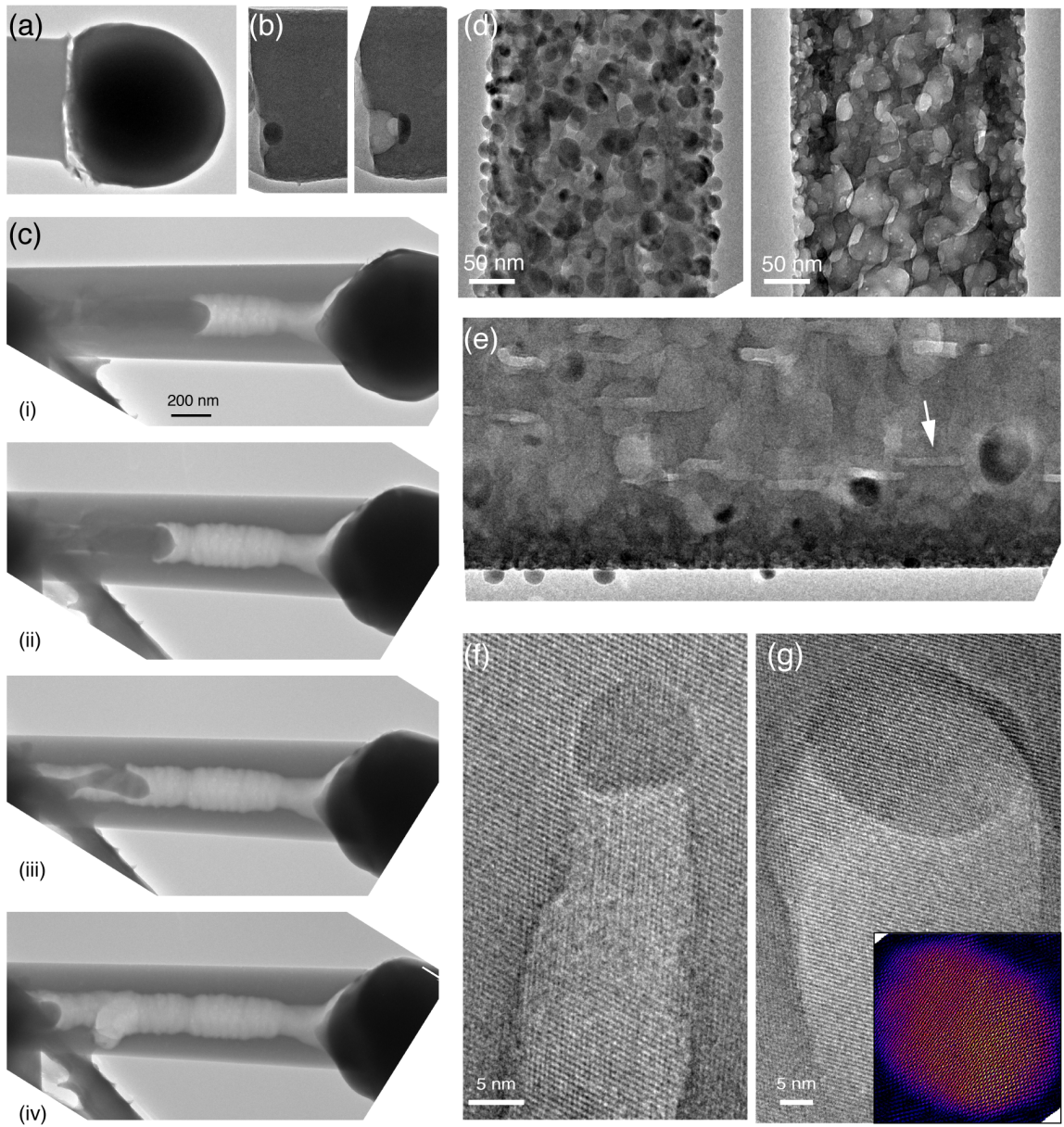

**Figure 4.** Ge-assisted etching of ZnO. (a) A gap at the Ge-ZnO interface developed after the sample was *in-situ* annealed at 650 °C under vacuum. (b) A residual catalyst particle creates a pit on the NW surface. (c) Behavior of a Ge rod within hollow core of a NT. (d) (left) Ge particles decorating a ZnO surface after digermane decomposition at 470 °C. (right) At higher temperature (610 °C), Ge particles dissolved into the ZnO, creating a porous structure. (e) Anisotropic etching creates long, faceted etched regions (arrow). (f,

g) Atomic-resolution TEM image showing the etching process. The particle removes ZnO materials at the moving front, leaving behind an empty trace of lower contrast. Inset of (g) is an inverse Fourier transform image of the particle displayed in false color (*Fire* in Look Up Table).

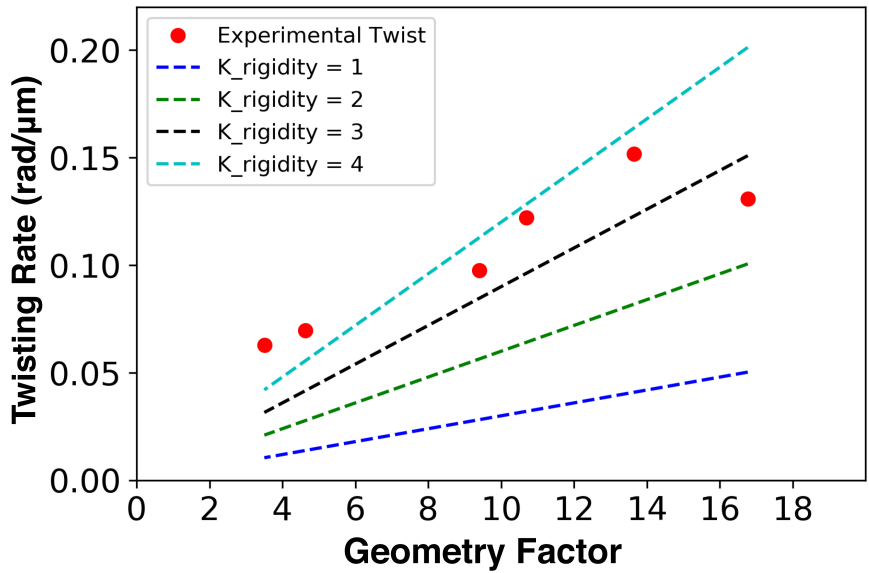

**Figure 5.** Twisting anomaly in etch-induced ZnO NTs. The measured twisting rates of NTs as the function of their geometry factor ($K_{Geometry}$) are displayed by the red dots. The calculated twisting rates $\alpha_{Tube}$ according to equation (5) for $K_{Rigidity}$ = 1, 2, 3 and 4 are depicted. The experimental values are bounded between $K_{Rigidity}$ = 3 and 4.

ToC figure

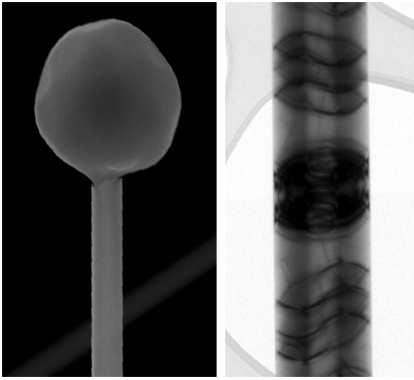